\documentclass[floatfix,prl,twocolumn,letterpaper,lengthcheck,superscriptaddress,showpacs,amssymb,amsmath,amsfonts,aps,altaffilletter,nofootinbib,nopreprintnumbers,longbibliography]{revtex4-1}

\usepackage{color}
\usepackage{hyperref}
\usepackage{amsmath}
\usepackage{amsthm}
\usepackage{multirow}
\usepackage{graphicx}
\usepackage{tikz}
\usetikzlibrary{calc}
\usepackage{placeins}
\usepackage{xspace}
\usepackage[normalem]{ulem}
\usepackage{physics,enumerate}

\usepackage{mathrsfs}
\DeclareSymbolFontAlphabet{\mathrsfs}{rsfs}

\newcommand{\lib}[1]{{\em #1\xspace}}

\newcommand{\Surf}{\ensuremath{\mathcal{S}}}

\def\outsubscript{\rm outer}
\newcommand{\Sout}{\Surf_{\outsubscript}}   % outer common MOTS S_outer
\newcommand{\Sonetwo}{\Surf_{1,2}}

\def\MOTSs{MOTSs\xspace}   % decide on plural form of "MOTS"
\begin{document}

\title[]{What Happens to Apparent Horizons in a Binary Black Hole Merger?}

\author{Daniel Pook-Kolb}
\affiliation{
    Max-Planck-Institut f\"ur Gravitationsphysik (Albert Einstein Institute),
    Callinstr. 38, 30167 Hannover, Germany
}
\affiliation{
    Leibniz Universit\"at Hannover, 30167 Hannover, Germany
}

\author{Robie A. Hennigar}
\affiliation{
    Department of Mathematics and Statistics, Memorial University,
    St. John's, Newfoundland and Labrador, A1C 5S7, Canada
}
\affiliation{
	Department of Physics and Astronomy, University of Waterloo, 
	Waterloo, Ontario, Canada, N2L 3G1
}
\affiliation{
	Department of Physics and Computer Science, Wilfrid Laurier University, 
	Waterloo, Ontario, Canada N2L 3C5
}

\author{Ivan Booth}
\affiliation{
    Department of Mathematics and Statistics, Memorial University,
    St. John's, Newfoundland and Labrador, A1C 5S7, Canada
}

% Keep explicit date, since otherwise arxiv will always insert the
% current date, which is useless and confusing
%\date{2020-11-23}

\begin{abstract}

We resolve the fate of the two original apparent horizons during the head-on merger of two non-spinning black holes.  We show that following the appearance of the outer common horizon and subsequent interpenetration of the original horizons, they continue to exist for a finite period of time before they are individually annihilated by unstable MOTSs.
The inner common horizon vanishes in a similar, though independent, way.
This completes the understanding of the analogue of the event horizon's 
``pair of pants'' diagram for the apparent horizon. Our result is facilitated by a new method for locating marginally outer trapped surfaces (MOTSs) based on a generalized shooting method.  We also discuss the role played by the MOTS stability operator in discerning which among a multitude of MOTSs should be considered as black hole boundaries. 
\end{abstract}

\maketitle

It is common practice to picture the merger of two black holes according to the famous ``pair of pants'' diagram, which describes the evolution of the event horizon during the merger. But what does the analogous picture for the apparent horizon (AH) look like? While the evolution of the event horizon has been understood for nearly half a century~\cite{hawking_ellis_1973, Matzner:1995ib}, the complete evolution of  apparent horizons during a merger has remained unresolved. 

%It is the purpose of the Letter to report on the resolution of this problem.

While the event horizon is well-suited to theoretical analyses, its teleological nature makes it less useful in highly dynamical or practical situations. In numerical simulations of mergers, marginally outer trapped 
surfaces (MOTSs) are used instead. A closed two-dimensional space-like surface $\mathcal{S}$ is a MOTS if light rays emitted normal to $\mathcal{S}$ are neither converging nor diverging in the outward direction. The outermost MOTS on a given slice is commonly called the apparent horizon. However, despite their importance as a quasi-local characterization of black holes, there remain many unresolved questions pertaining to the evolution of interior MOTSs during a merger.

While it is true that the details of what occurs within the event horizon (where all MOTSs are located) is causally disconnected from the rest of the universe, this does not mean that interior evolution is irrelevant. At the very least, it is conceptually important to understand to what extent MOTSs provide a physically reasonable description of the merger. Furthermore, the existence of a connected sequence of MOTSs between the initial and final states of the merger provides a means for  physical properties~\mbox{\cite{Dreyer:2002mx, Krishnan:2007va, Gupta:2018znn}} to be tracked throughout the full evolution. Finally, one may hope that there exist correlations between the dynamics in the strong field regime and properties of the distant spacetime. Indeed, such correlations have been shown to exist under certain circumstances~\cite{Rezzolla:2010df, Jaramillo:2011re, Jaramillo:2012rr, Gupta:2018znn, Prasad:2020xgr}.

The behaviour of AHs during the initial stages of a merger is well-known~\cite{hawking_ellis_1973}. Initially there are two individual AHs corresponding to two separate black holes. When these holes become sufficiently close to one another, a common AH forms surrounding the individual horizons, which continue to exist. This common horizon immediately splits into an inner and outer branch. The outer branch grows in area and becomes more symmetric, ultimately asymptoting to the event horizon. The inner common MOTS moves inward, becoming increasingly distorted.  

The bifurcation of the common horizon, combined with the fact that there are known exact solution examples of MOTSs weaving back and forth through time~\cite{BenDov:2004gh,Booth:2005ng}, led to the speculative idea that all MOTSs involved in the merger may be different components of a single world tube that weaves its way through time~\cite{Hayward:2000ca,bendov}. However, in most situations of interest MOTSs must be located numerically. Therefore, improvements in the understanding of the evolution of MOTSs during a merger have been in lockstep with improvements in the numerical methods used to locate them. For this reason, progress beyond this qualitative {speculation} was limited.

With the advent of more robust numerical finders for MOTSs~\cite{pook-kolb:2018igu}, it {is now}  possible to go beyond these initial stages of the merger and better understand the interior dynamics. The inner common MOTS continues to move inward and coincides with the
    union of the two AHs of the individual black holes at the
    moment these horizons touch. This can be interpreted as an instantaneous (non-smooth)
    merger of the three MOTSs~\cite{%
    PhysRevLett.123.171102,%
    PhysRevD.100.084044,%
    pook-kolb2020I,%
    pook-kolb2020II%
}. All continue to 
    exist beyond this point: the two individual horizons %then 
    interpenetrate and
    the inner common MOTS develops self-intersections. Identifying these  self-intersections was not possible with previous MOTS finders, as these had implicitly assumed that the MOTSs are `star-shaped'. {These} seemingly exotic surfaces have subsequently been shown to be rather generic. For example, there are an infinite number of self-intersecting MOTSs present within the horizon of the Schwarzschild black hole~\cite{Booth:2020qhb}. This raises the question of whether exotic MOTSs are {also} present {during a black hole} merger, and what role {they might play}.

In this Letter, we report three closely connected results. First, using a novel {MOTS}  finder, we identify for the first time an apparently infinite number of MOTSs present in {certain} Brill-Lindquist initial data. We then discuss the role that these new MOTSs play in resolving the final fate of the AHs of the two original black holes, finding that there do exist world tubes weaving back and forth through time. However, instead of a single smooth world tube, we find multiple distinct ones. Finally, we discuss the stability of these surfaces. The stability operator characterizes key merger processes and identifies which MOTSs should be thought of as black hole boundaries.

{\it A Multitude of MOTSs}. To locate MOTSs, we employ a novel shooting method. The procedure, while in the tradition of methods first developed in the 1970s \cite{CADEZ1974449}, is more versatile and applies to non-rotating though otherwise general  axisymmetric configurations~\cite{PaperI}. Implemented in \cite{pook_kolb_daniel_2021_4687700}, it 
can be applied to both analytically known initial data as well as to slices
obtained from numerical simulations. It overcomes a limitation of the method introduced in~\cite{pook-kolb:2018igu} as it does not require an initial guess for the shape of the surface to be located. As such,  it is ideally suited for locating {axisymmetric} MOTSs with geometries that are {both}  unexpected {and}  arbitrarily complicated.

The approach exploits axisymmetry to reduce the problem of locating a 2-surface to that of determining a curve $\gamma$, which we  {call} a \textit{MOTSodesic}. The full MOTS is the revolution of $\gamma$. Given an axisymmetric 3-surface \{$\Sigma, h_{ij}, D_i$\}, the {arclength parameterized} curve $\gamma(s)$ in the two-dimensional space orthogonal to the rotational Killing field $\varphi=\frac{\partial}{\partial\phi}$ is a MOTSodesic  if {it satisfies the 
two coupled second-order ODEs:}
\begin{equation}\label{motsodesic}
T^a D_a T^b = (N^c D_c (\ln R) + k_u) N^b  \equiv \kappa^+ N^b \, .
\end{equation}
Here $T^a$ is the unit-length tangent vector to $\gamma$, $N^a$ is its unit-length normal, $R$ is the circumferential radius, and $k_u$ is the trace of the extrinsic curvature of $\mathcal{S}$ with respect to the unit time-like normal $u$ to $\Sigma$.

%we introduce cylindrical-type coordinates $(\rho, z, \phi)$ where $\phi$ is the Killing coordinate associated with the axisymmetry and $\{\rho \ge 0 \, , -\infty < z < \infty \}$ cover the half-plane. A curve $\gamma \,  : \,  (\rho, z) = \left(P(s), Z(s)\right)$ is a MOTSodesic, i.e.

Henceforth, we restrict our considerations to the head-on merger of two non-spinning black holes. For this purpose, we use Brill-Lindquist (BL) initial data \cite{Brill:1963yv}. These describe a Cauchy slice $\Sigma$ which is time symmetric, i.e. with vanishing extrinsic curvature.
The three-metric is conformally flat,
$h_{ij} = \psi^4\delta_{ij}$, where $\delta_{ij}$ is the flat metric and the conformal factor is
\begin{equation}\label{eq:BLmetric}
    \psi = 1 + \frac{m_1}{2r_1} + \frac{m_2}{2r_2} \,,
\end{equation}
where $m_{1,2}$ are the bare masses of the black holes and $r_{1,2}$
are the (coordinate) distances to the respective punctures.

{Parametrizing $\gamma   :   (\rho, z) = \left(P(s), Z(s)\right)$ relative to cylindrical coordinates $(\rho, z, \phi)$,   the MOTSodesic equations  become}
\begin{align}
\ddot{P} &= \frac{\dot{Z}^2}{P} +  \frac{4\psi_\rho}{\psi^5} - \frac{6 \dot{P} ( \dot{P} \psi_\rho + \dot{Z} \psi_z)}{\psi}\,,
\\
\ddot{Z} &= -\frac{\dot{Z} \dot{P}}{P} +  \frac{4\psi_z}{\psi^5} - \frac{6 \dot{Z} ( \dot{P} \psi_\rho + \dot{Z} \psi_z )}{\psi}\,,
\end{align}
where subscripts denote partial derivatives and the arclength parameterization reads $ \psi^4(\dot{P}^2 + \dot{Z}^2) = 1$. 
The equations are solved in Mathematica using the shooting method (and the resulting MOTSs confirmed using the methods of~\cite{pook-kolb:2018igu}). 

\begin{figure}
\centering
\includegraphics[width=0.23\textwidth]{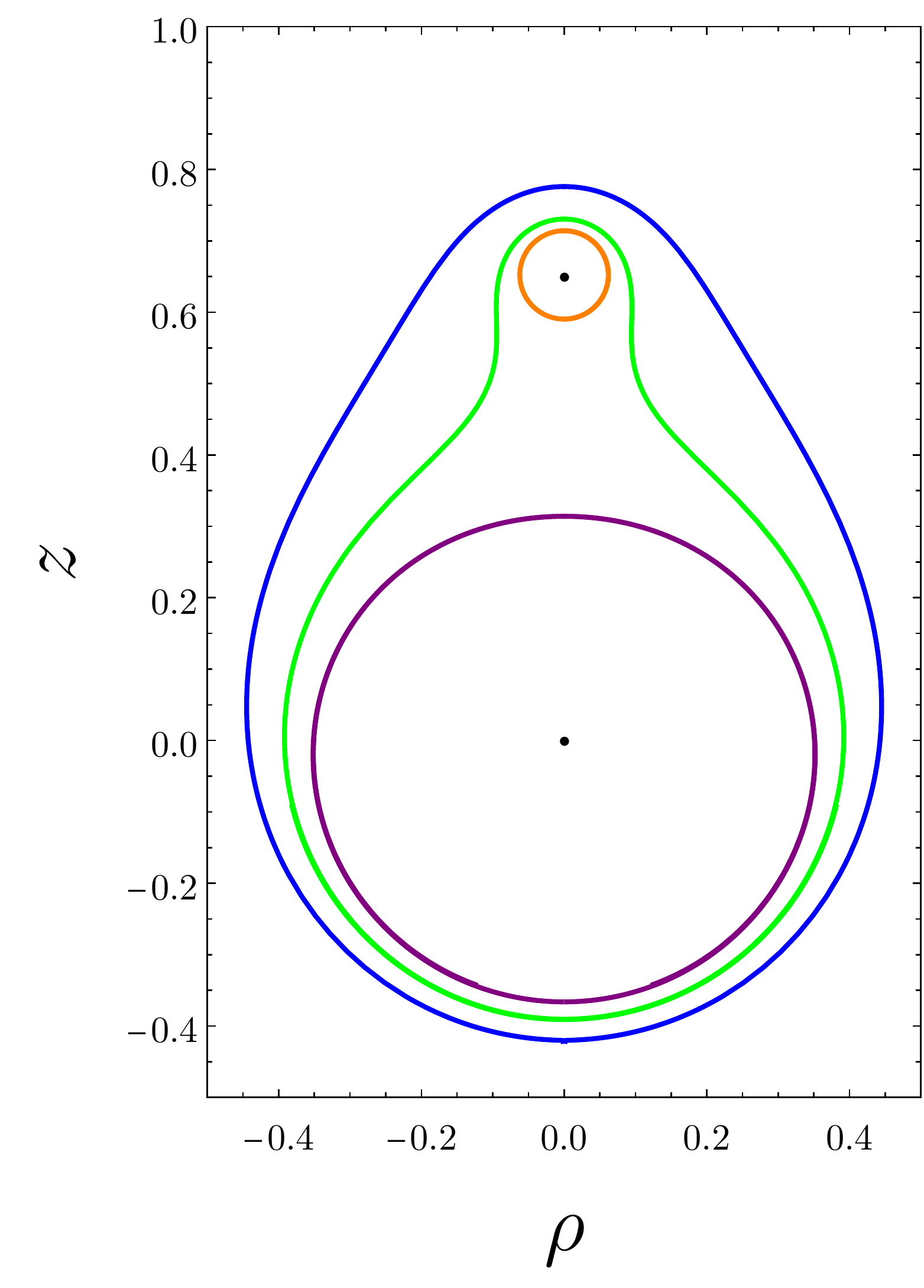}
\includegraphics[width=0.23\textwidth]{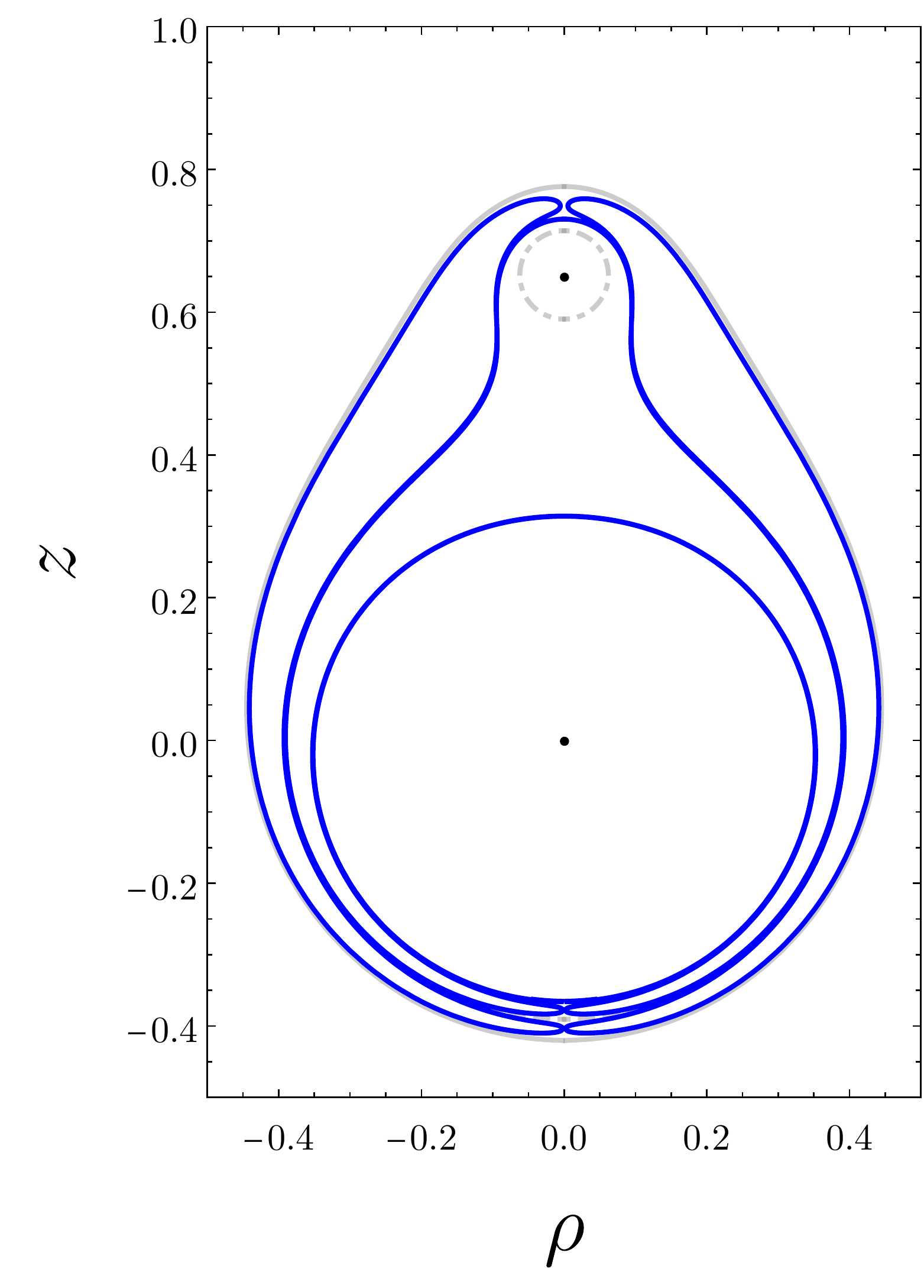}
\caption{MOTSs present in Brill-Lindquist initial data for $m_1 = 0.2$, $m_2 = 0.8$ and separation  $d = 0.65$. Left: The `standard' horizons consisting of the outer AH (blue), inner common MOTS (green), and AHs of individual black holes (orange and purple) reproduced using the shooting method. Right: An example of a new exotic MOTS.}
\label{MOTSmultitude}
\end{figure}

In addition to the  four `standard' MOTSs, we find a large number of more exotic MOTSs --- see Fig.~\ref{MOTSmultitude} for an example. 
In BL initial data, all new MOTSs are found between the outer AH and the two original AHs and can enclose either, both, or neither of the two punctures.
    These surfaces tend to closely `hug'  the outer and/or the individual AHs but can fold  a seemingly arbitrary number of times. By carefully tuning initial conditions, we find more and more MOTSs. The number is likely infinite. 

{\it The Fate of the Apparent Horizons}. The existence of new MOTSs in BL initial data raises a number of important questions. First, are these surfaces generic during a merger, or are they artefacts of the symmetry of the initial data? Second, if generic, what role (if any) do these surfaces play in the merger? Finally, with a seemingly infinite number of MOTSs present in a merger, how can one discern physically relevant surfaces that demarcate black hole boundaries?

To address these questions, we  {numerically evolve} the initial data. 
{Here we start}
with total ADM mass $M = m_1 + m_2 = 1$,  mass ratio
$q = m_2/m_1 = 2$ and distance parameter $d=0.9$,
corresponding to two black holes that are initially separate with no common AH or any of the new MOTSs present. To track  evolving MOTSs  we use the method described in~\cite{pook-kolb:2018igu,PhysRevD.100.084044} and available from
\cite{pook_kolb_daniel_2021_4687700}. 
{To locate new MOTSs we use not only the shooting method described earlier (detailed in~\cite{PaperI})
but also the ansatz that MOTSs only appear or disappear in pairs. When a MOTS cannot be tracked
into the future or past, we look for a ``nearby'' one with which it might annihilate or bifurcate.}

We perform our simulations with the
\lib{Einstein Toolkit} \cite{Loffler:2011ay,EinsteinToolkit:web}
and set up initial conditions using
\lib{TwoPunctures} \cite{Ansorg:2004ds}.
The Einstein equations are evolved in the BSSN formulation using an
axisymmetric version of
\lib{McLachlan} \cite{Brown:2008sb}, which
uses \lib{Kranc} \cite{Husa:2004ip,Kranc:web}
to generate C++ code.
We work with a
$1+\log$ slicing and a $\Gamma$-driver shift condition
\cite{Alcubierre:2000xu,Alcubierre:2002kk}.
Most  results are obtained with a spatial grid resolution
of $1/\Delta x = 720$.
Lower resolutions and shorter simulations
with resolutions up to $1/\Delta x = 1920$
are used to assess convergence and
resolve certain features (see \cite{PaperII, PhysRevD.100.084044} for more details).

\begin{figure}
\includegraphics[width=0.9\linewidth]{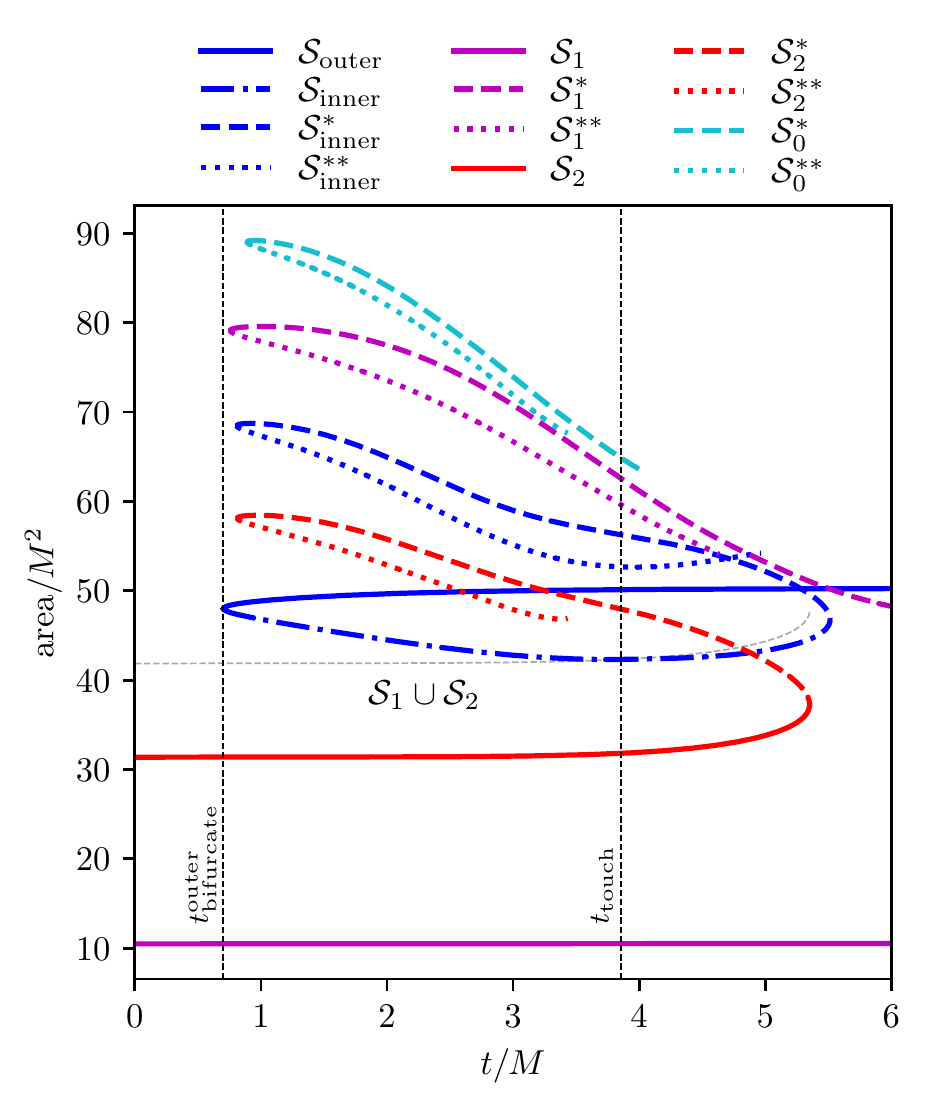}
\caption{The area of the various MOTSs as a function of time. Lines of the same colour {indicate} continuous world tubes moving forward and backward in time. The {pale gray} dashed line indicates the sum of areas of $\mathcal{S}_{1,2}$. For numerical reasons, we lose track of some of the MOTSs --- this is {why some of the curves abruptly terminate}.}
\label{pants}
\end{figure}

\begin{figure}
\includegraphics[width=0.9\linewidth]{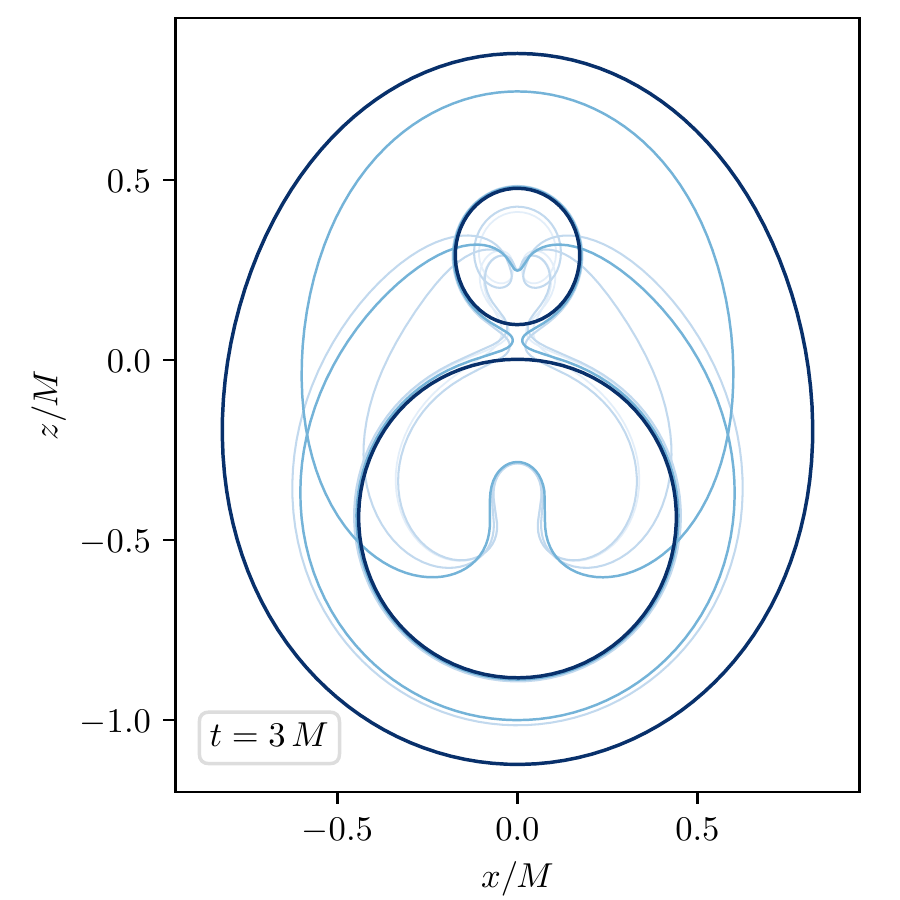}
\caption{A ``snapshot'' of Fig.~\ref{pants} showing a variety of MOTSs. The three dark lines correspond to $\Sout$ and $\Sonetwo$. Two of the shown surfaces exhibit self-intersections. Lighter colors indicate a larger number of negative eigenvalues of the stability operator (see below and \cite{PaperI, PaperII} for details). Note that, in contrast to BL initial  data, the absence of time symmetry means that it is now
possible for the new MOTS to penetrate the inner AHs. }
\label{pantsSlice}
\end{figure}

Our main results are illustrated in Fig.~\ref{pants}, which shows the {area} evolution of several relevant MOTSs, and Fig.~\ref{pantsSlice} which shows 
 {those} MOTSs at a particular moment of time:  $\mathcal{S}_{\rm outer}$ is the common AH, $\mathcal{S}_{\rm inner}$ is the inner common MOTS, and $\mathcal{S}_{1,2}$ correspond to the AHs of the individual black holes.  {Beyond} these standard MOTSs, we find many
new MOTSs;  a selection {appear} in Fig.~\ref{pants}. 
{ The new surfaces all form through bifurcations, splitting into outer and inner branches. Hence exotic MOTSs are not solely artefacts of time symmetry. }
The shown MOTSs all form \textit{after} the outer AH has formed, and despite several having larger area than $\mathcal{S}_{\rm outer}$, are all  contained {within} it.

\begin{figure*}
\includegraphics[width=0.3\textwidth]{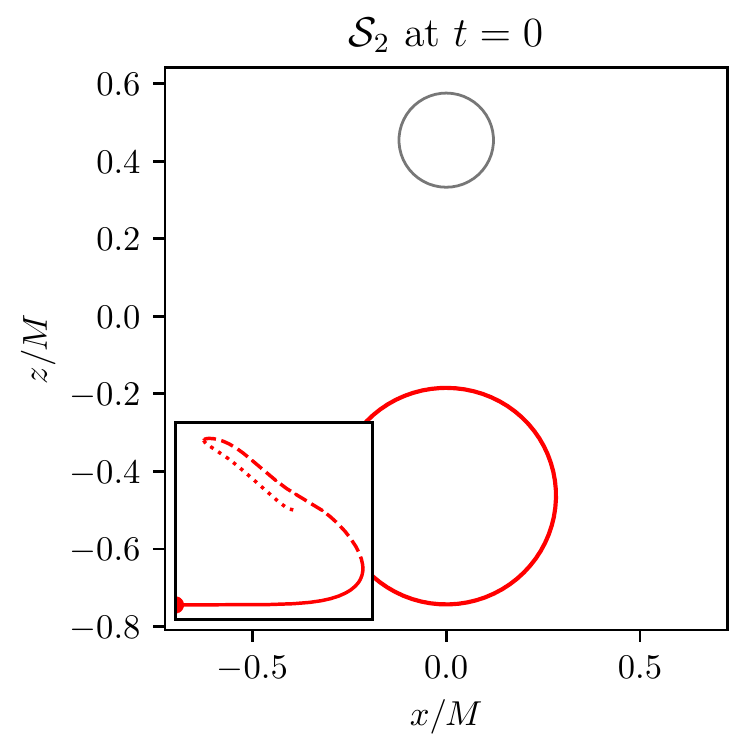}
\includegraphics[width=0.3\textwidth]{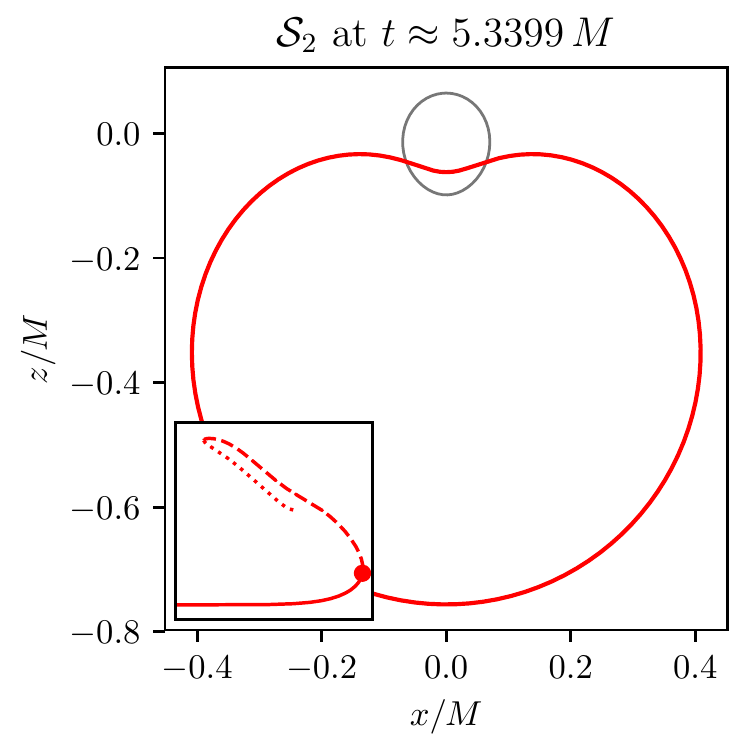}
\includegraphics[width=0.3\textwidth]{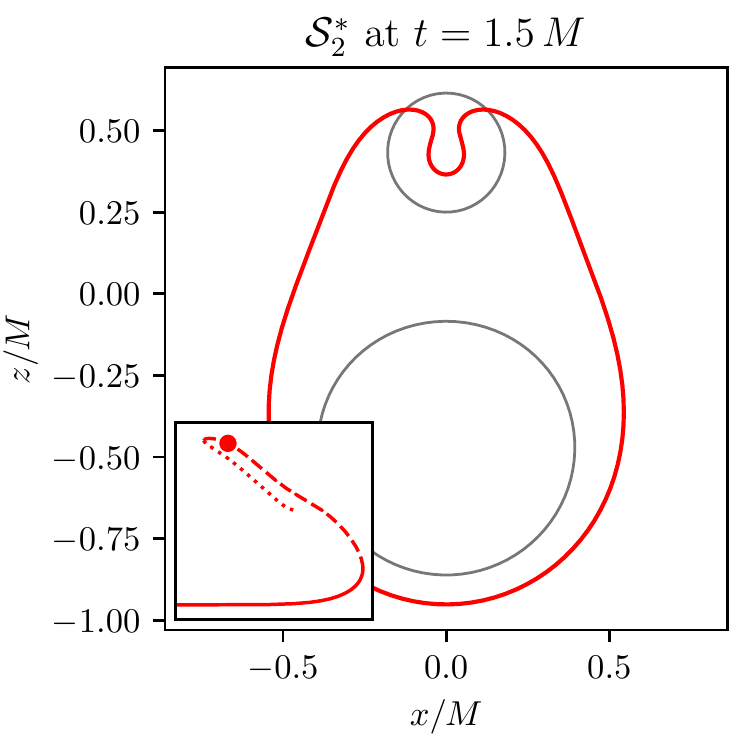}
\caption{The annihilation of $\mathcal{S}_2$. The inset shows, with a red dot, where along the world tube the shown MOTS occurs [see Fig.~\ref{pants}].}
\label{deathOfS1}
\end{figure*}

The figure  makes clear that exotic MOTSs are essential to understanding the final fate of the AHs of the individual black holes.  Both $\mathcal{S}_{\rm inner}$ and $\mathcal{S}_2$ are independently annihilated by new MOTSs. We have good indications that $\mathcal{S}_1$ is annihilated by $\mathcal{S}_1^*$, but this could not be fully resolved in our simulation due to the MOTSs becoming too close to the punctures, where the resolution is necessarily worse. Nonetheless, it seems clear that {our} new MOTSs provide a mechanism by which the AHs of the original black holes are annihilated.   To illustrate this in greater detail, we present in Fig.~\ref{deathOfS1} several ``snapshots'' of the evolution of  $\mathcal{S}_2$.

\begin{figure}
    \includegraphics[width=0.9\linewidth]{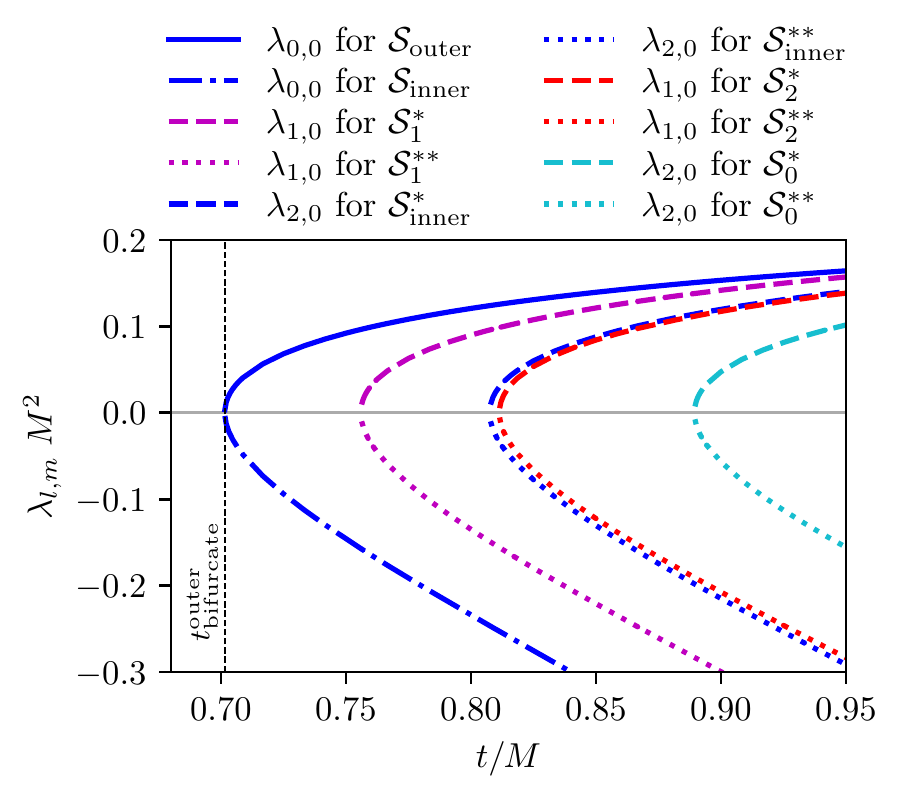}%
    \caption{\label{fig:stability_bifurcations}%
        Eigenvalues (with $m=0$) of the stability operator for the ten \MOTSs participating in the
        five bifurcations  shown in Fig.~\ref{pants}.
        For each MOTS, we show the respective eigenvalue which tends to
        zero.
    }
\end{figure}

{\it The Role of Stability}. The stability operator is key in understanding the details of the picture so far described. For a non-spinning axisymmetric MOTS in vacuum,   {this} takes the form~\cite{PaperI}:
\begin{equation}\label{stabilityOperator}
L_{\Sigma} \Psi = - \Delta_\mathcal{S} \Psi + \left(\frac{\mathcal{R}}{2} - 2 |\sigma_+|^2 \right) \Psi\,,
\end{equation}
where $\Delta_\mathcal{S}$ is the Laplacian on $\mathcal{S}$, $\mathcal{R}$ is its scalar curvature, and $\sigma_+^{AB}$ is the {shear in the outward null direction}. Geometrically this operator arises when one considers local deformations of a given MOTS, {with} the function $\Psi$, fully determining the deformation~\cite{Andersson:2005gq}. Intuitively $\Psi$  can be thought of as {measuring} the {normal} distance between
the MOTS and its deformation. 

Of particular interest is the eigenvalue spectrum $L_\Sigma \Psi = \lambda_{l, m} \Psi$, where the
$m=0$ eigenvalues correspond to axisymmetric deformations and in our non-spinning, axisymmetric case, all eigenvalues are real.
$\lambda_{l, m}$ encodes information about geometry of the MOTS. Especially important is the principal (smallest) eigenvalue $\lambda_{0,0}$~\cite{Andersson:2005gq, Andersson:2007fh, Andersson:2008up}. If  $\lambda_{0,0}$ is positive, then the MOTS is called strictly stable and evolves smoothly into the future. The world tubes traced out by such MOTSs are everywhere expanding  and space-like.  Furthermore, these MOTSs act as barriers between  trapped and untrapped surfaces in their vicinity. The vanishing of the principal eigenvalue has been identified with the bifurcation of $\mathcal{S}_{\rm outer}/\mathcal{S}_{\rm inner}$\cite{pook-kolb:2018igu,PhysRevD.100.084044,pook-kolb2020II}.

We find these general considerations borne out in our work, {though} with a number of important  {modifications} and caveats. First, of the potentially infinite number of MOTSs present in the interior of the merger, only three are strictly stable: $\mathcal{S}_{\rm outer}$, $\mathcal{S}_1$ and $\mathcal{S}_2$. Only these surfaces act as barriers
%\footnote{\robie{\sout{This is in the sense of \cite{Andersson:2007fh}, i.e. every small outward (inward) deformation leads to an outer untrapped (trapped) surface. In this perturbative sense, it is a local barrier for MOTSs in some neighbourhood.
% However, as is clear from FIG.\ref{pantsSlice}, the inner stable MOTSs 
% can still be penetrated by non-perturbative MOTSs that are not entirely contained in that neighbourhood.
%}}},
 and they are also the only ones to have everywhere expanding space-like world tubes. These properties are precisely what one would associate with horizons. Therefore stability provides an unambiguous criterion by which the MOTSs corresponding to black hole boundaries may be identified.  {This observation} underlies our choice of terminology for the AH advocated in this work: an AH is a stable MOTS ($\lambda_{0,0} \geq 0$, i.e. we include the marginal case). All other MOTSs located have at least one negative eigenvalue. 
%In fact, the picture that has emerged can be understood rather simply: each time a given world tube turns around in time, the number of negative eigenvalues increases.

We find that, associated to every bifurcation/annihilation is the vanishing of an eigenvalue of the stability operator --- see Fig.~\ref{fig:stability_bifurcations} for the case of the bifurcations. The formation/annihilation of the AHs coincide with the vanishing of the principal eigenvalue, in line with the results described above. However, a key finding of our work is that, in the case that two MOTSs are unstable, it is the vanishing of one of the \textit{higher} $m=0$ eigenvalues that coincides with their bifurcation/annihilation. A simple picture emerges: the MOTSs foliating a given world tube accrue an additional  negative $m = 0$ eigenvalue with each fold in time.

Finally, the properties of the spectrum provide further evidence for the evolution we have advocated in this letter. At a given bifurcation/annihilation event, we find that the numerical values of the eigenvalues of the surfaces involved connect smoothly (see again Fig.~\ref{fig:stability_bifurcations}). This provides robust numerical evidence for the world tubes to be locally smooth across the bifurcations/annihilations shown in Fig.~\ref{pants}.

{\it Summary}. We have shown that the interior of a black hole merger is far richer than previously thought, containing a large (possibly infinite) number of hitherto unidentified MOTSs. These MOTSs were initially located using a new shooting method that
% removes 
sidesteps the drawback in existing finders of requiring an initial guess for the surface of interest. The additional surfaces play a crucial role in the interior dynamics of the merger, and are responsible for the annihilation of the AHs of the original black holes. As such with these new MOTSs we reveal, for the first time, the full story of how two black holes become one, giving the analogue of the ``pair of pants'' diagram for the AH. The picture is considerably more complex than the equivalent picture for the event horizon and involves several world tubes that weave  back and forth in time. The stability of MOTSs has played a clarifying role in our work. Rather than obscuring the utility of the quasi-local horizon framework, the multitude of MOTSs present during the merger actually highlights the rarity of stable MOTSs. Of all the MOTSs we have located, only three are stable, and these are precisely those that are most naturally associated with black hole boundaries. Moreover, we have found that associated to each bifurcation/annihilation event is the vanishing of \textit{some} eigenvalue of the stability operator, not usually the principal one. This observation may {aid in the }  analytical understanding of the world tubes traced out by unstable MOTSs.

%\robie{
%A natural question is to what extent our results apply to situations lacking axisymmetry. 
%}

%The picture that emerges is similar in spirit, but considerably more complicated, than speculative ideas that the apparent horizons may be components of a single world tube weaving through time. The reality is more complicated than this, but we do observe world tubes weaving back and forth in time.  

\begin{acknowledgements}
     {\it Acknowledgements}. We would like to express our gratitude to 
    Graham Cox, 
    Jose~Luis~Jaramillo,
    Badri~Krishnan, Hari Kunduri and the members of the 
    Memorial University Gravity Journal Club
    for valuable discussions and suggestions. IB was supported by the Natural Science and Engineering Research Council of Canada Discovery Grant 2018-0473. 
   The work of RAH was supported by the Natural Science and Engineering Research Council 
   of Canada through the Banting Postdoctoral Fellowship program and also by AOARD Grant FA2386-19-1-4077.
\end{acknowledgements}

\bibliography{blmotos}{}

\end{document}